\documentstyle[12pt]{article}
\begin{document}
\rightline{UT-721 ,\ 95}
\large
\noindent
Phase operator for the photon, an index theorem, and quantum anomaly{%
\renewcommand{\thefootnote}{\fnsymbol{footnote}}
\footnote[2]{To be published in the Proceedings of ISQM-95,Tokyo
 (Elsevier Pub., Amsterdam)}
}%
\\

\normalsize
\noindent
Kazuo Fujikawa
\\

\noindent
Department of Physics, University of Tokyo\\
Bunkyo-ku, Tokyo 113, Japan
\\
\\

An index relation $dim\ ker\ a - dim\ ker\ a^{\dagger} = 1$ is
satisfied by the creation and annihilation operators $a^{\dagger}$
and $a$ of a harmonic oscillator. Implications of this analytic index on the
possible form of the phase operator are discussed. A close
analogy between the present phase operator problem and chiral anomaly in gauge
theory, which is associated with Atiyah-Singer index
theorem, is emphasized.\\
\\
\nopagebreak
1. INDEX RELATIONS\\
\par
The quantum phase operator has been studied by various authors in the
past[1-4]. We here remark on the absence of a hermitian
phase operator and the lack of a mathematical basis for
$\Delta N\Delta\phi \geq 1/2$,  on the basis of a notion of
index or an index theorem. To be specific,
we study the simplest one-dimensional harmonic oscillator defined by
 $h =  a^{\dagger}a + 1/2$
where $a$ and $a^{\dagger}$ stand for the annihilation and creation
operators satisfying  the standard commutator $[a, a^{\dagger} ] = 1
$.
The vacuum state $|0\rangle$
is annihilated by $a$, $ a|0 \rangle = 0$
which ensures the absence of  states with negative norm.
The number operator defined by $N = a^{\dagger}a$
then has non-negative integers as eigenvalues, and the annihilation
operator
$a$ is represented by
\begin{equation}
a  = |0\rangle \langle 1|+ |1\rangle \langle 2|\sqrt{2}
              + |2\rangle \langle 3|\sqrt{3} + ....
\end{equation}
in terms of the eigenstates $|k\rangle$ of the number operator,
$N|k\rangle = k|k\rangle$,
with $k = 0, 1, 2, ... $. The creation operator $a^{\dagger}$ is given by the
hermitian conjugate of $a$ in (1).

In the representation of $a$ and $a^{\dagger}$ specified above we have the
index condition
\begin{equation}
dim\ ker\ a - dim\ ker\ a^{\dagger} = 1
\end{equation}
where $dim\ ker\ a$,for example, stands for the number of
normalizable basis vectors $u_{n}$ which satisfy $au_{n}=0$.

In the conventional notation of index theory, the relation (2) is
written by using the trace of well-defined operators as
\begin{equation}
Tr(e^{-a^{\dagger}a/M^{2}}) - Tr(e^{-aa^{\dagger}/M^{2}}) = 1
\end{equation}
The relation (3) is confirmed for the standard representation (1) as
,$1 + (\sum_{n=1}^{\infty}e^{-n/M^{2}}) - (\sum_{n=1}^{\infty}e^{-n/M^{2}})=
1$,
independently of the value of $M^{2}$:Eq.(3) may be taken as an altertnative
definition of the index.

If one should suppose the existence of a well defined hermitian phase operator
$\phi$, one would have a polar decomposition
\begin{equation}
a = U(\phi)H = e^{i\phi}H
\end{equation}
as was originally suggested by Dirac[1]. Here  $U$ and $H$
stand for unitary and hermitian operators, respectively.
If (4) should be valid, we have in the same notation as (3)
\begin{equation}
Tr(e^{-a^{\dagger}a/M^{2}}) - Tr(e^{-aa^{\dagger}/M^{2}})
 = Tr(e^{-H^{2}/M^{2}}) - Tr(e^{-UH^{2}U^{\dagger}/M^{2}})= 0
\end{equation}
This relation when combined with (3) constitutes a proof of the absence of
a hermitian phase operator in the framework of index theory.

The basic utility of the notion of index or an index theorem lies
in the fact that the index as such is an integer and remains invariant
under a wide class of continuous deformation.
As an example, the unitary time development of $a$ and $a^{\dagger}$ dictated
by the Heisenberg equation of motion, which includes a
fundamental phenomenon such as squeezing, does not alter the index
relation.

If one truncates the representation of $a$ to any finite dimension,
for example, to an $(s + 1)$ dimension with $s$ a positive
integer, one obtains by noting $a_{s}^{\dagger}|s\rangle =0$
\begin{equation}
dim\ ker\ a_{s} - dim\ ker\ a_{s}^{\dagger} = 0
\end{equation}
where $a_{s}$ stands for an $s+1$ dimensional truncation of $a$.
This relation (6) is proved  by noting that  non-vanishing eigenvalues of
$ a_{s}^{\dagger}a_{s}$ and  $  a_{s}a_{s}^{\dagger}$ are in one-to-one
correspondence:
In the eigenvalue equations
\begin{equation}
a_{s}^{\dagger}a_{s}u_{n} = \lambda^{2}_{n}u_{n}
\end{equation}
one may define $v_{n} = a_{s}u_{n}/\lambda_{n}$ for $\lambda_{n} \neq 0$.
One then obtains
$a_{s}a_{s}^{\dagger}v_{n} = \lambda_{n}^{2}v_{n}$
by multiplying $a_{s}$ to  both hand sides of eq. (7). For any finite
dimensional matrix , one also has
$Tr(a_{s}^{\dagger}a_{s}) = Tr(a_{s}a_{s}^{\dagger})$.
These two facts combined then lead to the statement that
$a_{s}^{\dagger}a_{s}$ and $a_{s}a_{s}^{\dagger}$ contain the same number of
zero eigenvalues, which implies the relation (6) or
\begin{equation}
 Tr_{(s+1)}(e^{-a^{\dagger}_{s}a_{s}/M^{2}})
-Tr_{(s+1)}(e^{- a_{s}a^{\dagger}_{s}/M^{2}})=0
\end{equation}
where $Tr_{(s+1)}$ denotes an $s+1$ dimensional trace.
This index relation holds independently
of $M^{2}$ and $s$.\\
\\
\noindent
2. PHASE OPERATORS\\
\\
{}From the above analysis of the index condition , the polar
decomposition (4) could be consistently
defined if one truncates the dimension of the representation
space to a finite $s+1$ dimension . This is the approach adopted by Pegg
and Barnett[3] in their definition of a hermitian phase operator
\begin{equation}
e^{i\phi}=|0\rangle \langle 1| + |1\rangle \langle 2| + ...
          + |s-1\rangle \langle s|
          + e^{i(s+1)\phi_{0}}|s\rangle \langle 0|
\end{equation}
where $\phi_{0}$ is a constant c-number. One can also define hermitian
functions
\begin{equation}
\cos\phi = \frac{1}{2} (e^{i\phi} + e^{-i\phi}),\
\sin\phi = \frac{1}{2i}(e^{i\phi} - e^{-i\phi})
\end{equation}
with $e^{-i\phi} = (e^{i\phi})^{\dagger}$.These operators satisfy , among
others,
\begin{equation}
{[} \cos\phi, \sin\phi] = 0,\ \cos^{2}\phi + \sin^{2}\phi = 1
\end{equation}
Their basic
idea is then to let $s$ arbitrarily large later.
The kernels of  $a_{s}$ and $a_{s}^{\dagger}$ are given by
\begin{equation}
 ker\ a_{s} = \{ |0\rangle\}, \ \ ker\ a_{s}^{\dagger} = \{ |s\rangle\}
\end{equation}
 The limit
$s \rightarrow \infty$ of the relation (6) is thus singular since the kernels
in (12) are ill-defined for $s \rightarrow \infty$.

On the other hand, a non-hermitian phase operator which faithfully
reflects the index condition (2) was introduced by Susskind and
Glogower[2]
\begin{equation}
e^{i\varphi}=\frac{1}{\sqrt{N+1}}a = |0\rangle \langle 1|+ |1\rangle \langle 2|
+ |2\rangle \langle 3| + ....
\end{equation}
with $dim\ ker\ e^{i\varphi} - dim\ ker\ (e^{i\varphi})^{\dagger}=1$.
This $e^{i\varphi}$ is not unitary, as is witnessed by
$(e^{i\varphi})^{\dagger}e^{i\varphi}= 1 - |0\rangle\langle 0|$ and
$e^{i\varphi}(e^{i\varphi})^{\dagger}= 1$.
One may still define hermitian operators
\begin{equation}
C(\varphi)= \frac{1}{2}(e^{i\varphi} + (e^{i\varphi})^{\dagger}),\
S(\varphi)= \frac{1}{2i}(e^{i\varphi} - (e^{i\varphi})^{\dagger})
\end{equation}
which, together with the number operator $N$, satisfy
\begin{displaymath}
{[} N  , C(\varphi)]      =  -i S(\varphi),\
{[} N  , S(\varphi)]       =   i C(\varphi)
\end{displaymath}
\begin{equation}
{[}C(\varphi) , S(\varphi)]   = \frac{1}{2i}|0\rangle \langle 0|,\
C(\varphi)^{2} + S(\varphi)^{2} = 1 - \frac{1}{2}|0\rangle \langle 0|
\end{equation}
The last two relations in (15) are ``anomalous'' whereas the relations in (11)
are normal from a view point of classical-quantum analogy.

It was shown in ref. (5) that the apparently ``anomalous'' relations
(15) are in fact more consistent with quantum phenomena in the sense
that one can prove the uncertainty relations
\begin{displaymath}
\Delta N \Delta \sin\phi \geq \Delta N \Delta S(\varphi) \geq
\frac{1}{2}|\langle p|C(\varphi)|p\rangle| =   \frac{1}{2}|\langle
p|\cos\phi|p\rangle|,
\end{displaymath}
\begin{equation}
\Delta \cos\phi \Delta \sin\phi \geq \Delta C(\varphi) \Delta S(\varphi) \geq
\frac{1}{4}|\langle p|0\rangle \langle 0|p\rangle|\geq 0
\end{equation}
for $any$ physical state $|p\rangle$ which satisfy $\langle p|N^{2}|p
\rangle <  \infty$.
The uncertainty for the relations in (11) is $always$ larger than
the uncertainty for the relations in (15). If a state $|p\rangle$ is a minimum
uncertainty state for the set of variables $(N, \cos\phi,
\sin\phi)$ in the sense that it gives rise to an equality in the uncertainty
relation, the same state $|p\rangle$ is automatically the minimum uncertainty
state for the set of variables $(N, C(\varphi),
S(\varphi))$. But the other way around is not necessarily true in
general. This discrepancy is caused by the state $|s\rangle$ in (9), which is
also responsible for the index relation (12). The unitary operator $e^{i\phi}$
in (9)  always carries a vanishing index
\begin{equation}
dim\ ker\ e^{i\phi} - dim\ ker\ (e^{i\phi})^{\dagger} = 0
\end{equation}
since the unitary operators simply relabel the basis vectors without
changing the number of them. This index mismatch of (17) with (13),which is
caused by the state $|s\rangle$, thus leads to a
 substantial deviation from minimum uncertainty in a characteristically quantum
domain with small average photon numbers[5].The deviation from the minimum
uncertainty becomes appreciable in a characteristically quantum domain since
the non-vanishing matrix element, for example, $|\langle
s|\sin\phi|p\rangle|^{2}$, for whatever large $s$ may be, becomes more
appreciable compared with other states $|n\rangle$ contributing to the matrix
elements
$\sum|\langle n|\sin\phi|p\rangle|^{2}$.\\
\\
\noindent
3. ANALOGY WITH CHIRAL ANOMALY\\
\\
It was also emphasized in ref. (5) that this kind of phenomenon, i.e., an
apprently anomalous behavior (15) is in fact more consistent with quantum
phenomena is a close analogy of chiral anomaly in gauge theory,
which is related to the Atiyah-Singer index theorem similar to (2).
In the analysis of chiral anomaly in gauge theory, the mass of the
Pauli-Villars regulator provides a cut-off parameter and one obtains a normal
result expected on the basis of classical-quantum analogy
 for any finite cut-off parameter. But in the limit of the large
regulator mass, the cut-off parameter does not decouple and induces
an anomalous term which is dictated by the Atiyah-Singer index theorem.

Precisely the same phenomenon takes place in the phase operator problem if one
identifies the parameter $s$ as the cut-off parameter. One may rewrite the
relation (8) as
\begin{equation}
 Tr_{(s+1)}(e^{-a^{\dagger}_{s}a_{s}/M^{2}})
-Tr_{(s)}(e^{- a_{s}a^{\dagger}_{s}/M^{2}}) =
Tr_{(s+1)}(e^{-a_{s}a^{\dagger}_{s}/M^{2}})
-Tr_{(s)}(e^{- a_{s}a^{\dagger}_{s}/M^{2}}) = 1
\end{equation}
where $Tr_{(s)}$ stands for the trace over the first $s$ dimensional
subspace of the $s+1$ dimensional space, and the right-hand side of
(18) is the contribution of the state $|s\rangle$, i.e., the cut-off parameter.
It can be confirmed that each term in the left-hand side
of (18) has a well-defined limit for $s \rightarrow \infty$ and one recovers
the non-trivial index relation (2), which in turn prohibits
the existence of a hermitian phase operator. On the basis of this
analysis, it was suggested in ref. (5) that the anomalous behavior in (15) is
an inevitable and unavoidable quantum effect, not an artifact of our
insufficient definition of phase operator.\\
\\
\noindent
REFERENCES\\
\\
\noindent
1. P. A. M. Dirac, Proc. Roy. Soc. (London) A114 (1927) 243.\\
2. L. Susskind and J. Glogower, Physics 1 (1964) 49.\\
3. D. T. Pegg and S. M. Barnett, Phys. Rev. A39 (1989) 1665.\\
4. P. Carruthers and M. M. Nieto, Rev. Mod. Phy. 40 (1968) 411 and references
therein.\\
5. K. Fujikawa, to be published in Phys. Rev. A;hep-th/9411066.

\end{document}